\documentclass[12pt,a4paper]{article}
\usepackage[english]{babel}
\usepackage[numbers,comma,square,compress]{natbib}
\usepackage{amsmath,amssymb,amsfonts}
\usepackage{ifpdf}
\ifpdf  % Compiling PDF with pdflatex
  \usepackage[pdftex,a4paper,margin=26mm]{geometry}
\else  % Compiling DVI with latex
  \usepackage[dvips,a4paper,margin=26mm]{geometry}
\fi

\def\mM{\mathcal{M}}
\def\mV{\mathcal{V}}
\def\mL{\mathcal{L}}
\def\tg{\tilde{g}}
\def\tnabla{\tilde{\nabla}}
\def\hg{\hat{g}}
\def\bn{\mathbf{n}}

\def\mH{\mathcal{H}}
\def\mG{\mathcal{G}}
\newcommand{\sR}{{}^{(3)}\!R}
\newcommand{\Kt}{{}^{(2)}\!K}
\newcommand{\email}[1]{\href{mailto:#1}{#1}}
\providecommand{\href}[2]{#2}

\begin{document}
\begin{center}
{\Large Can TeVeS be a viable theory of gravity?}

\vspace{1em}
Masud Chaichian$^a$,
Josef Kluso\v{n}$^b$,
Markku Oksanen$^a$,
Anca Tureanu$^{a,}$\footnote{Email addresses:
\email{masud.chaichian@helsinki.fi} (M. Chaichian),
\email{klu@physics.muni.cz} (J. Kluso\v{n}),
\email{markku.oksanen@helsinki.fi} (M. Oksanen),
\email{anca.tureanu@helsinki.fi} (A. Tureanu)}\\
\vspace{1em}
$^a$\textit{Department of Physics, University of Helsinki, P.O.
Box
64,\\ FI-00014 Helsinki, Finland}\\
\vspace{.3em}
$^b$\textit{Department of Theoretical Physics and Astrophysics, Faculty
of Science,\\
Masaryk University, Kotl\'a\v{r}sk\'a 2, 611 37, Brno, Czech Republic}\\
\end{center}

\begin{abstract}
Among modified gravitational theories, the Tensor-Vector-Scalar (TeVeS)
occupies a special place -- it is a covariant theory of gravity that
produces the modified Newtonian dynamics (MOND) in the nonrelativistic
weak field limit and explains the astrophysical data at scales larger
than that of the Solar System, without the need of an excessive amount
of invisible matter. We show that, in contrast to other modified
theories, TeVeS is free from ghosts. These achievements make TeVeS (and
its nonrelativistic limit) a viable theory of gravity. A speculative
outlook on the emergence of TeVeS from a quantum theory is presented.
\end{abstract}

\vspace{.5em}
\noindent
{\small Keywords: tensor-vector-scalar gravity, modified Newtonian
dynamics, Hamiltonian formalism,\\
\hspace*{4.7em} ghost instability, quantum gravity, dark matter}

% PACS numbers
\noindent
{\small PACS: 04.50.Kd, 04.20.Fy, 04.60.-m, 95.35.+d}

\section{Introduction}
The current accepted theory of gravity is Einstein's General Relativity
(GR), which has been experimentally tested in the Solar System with
great success. On the galactic and cosmological scales, however, the
observed dynamics does not agree with the observed distribution of
matter, when GR is taken as the theory of gravity.
In order to make GR consistent with the observations on the
galactic and cosmological scales, we have to postulate new invisible
forms of energy, commonly referred to as dark matter and dark energy,
which constitute the major part of the energy in the Universe. Neither
of these dark elements has been observed by means other than their
interaction with gravity. Since the postulation of such invisible
elements may be a specious solution, we have to consider some other
alternatives: GR may have to be amended.

We consider the Tensor-Vector-Scalar theory of gravity
\cite{Bekenstein:2004ne} (TeVeS) as an alternative to GR.
TeVeS is a relativistic theory of gravity, which produces
the modified Newtonian dynamics \cite{Milgrom:1983ca,Bekenstein:1984tv}
(MOND) in the nonrelativistic weak field limit.
According to the MOND paradigm, there exists an acceleration scale
$a_0$ such that for accelerations smaller than $a_0$, Newton's second
law is modified so that the gravitational force is proportional to
the square of the particle's acceleration.
It is remarkable that this simple proposal is so successful in
explaining the galactic rotation curves \cite{Milgrom:2004ba},
thus alleviating the need for dark matter on the galactic scales.
For a review on MOND and TeVeS, see \cite{Famaey:2012lrr}
and \cite{Bekenstein:2012ny}, respectively. Further on,
TeVeS has been shown to be free of acausal propagation of perturbations
and it is in agreement with solar system tests \cite{Bekenstein:2004ne}.

A priori, the missing mass could be composed of baryons in objects other
than stars, such as brown dwarfs, Jupiter sized planets, or any kind of
normal matter which is unseen presently. However, the experimental
observations do not confirm the abundance of these objects
\cite{Afonso:2002xq}. The simplest explanation for the discrepancy
between the dynamics and matter distribution is to postulate a new form
of non-baryonic matter, the so-called dark matter, which does not
interact with electromagnetic radiation. In order to explain the
observed extra gravitational force, the abundance of dark matter has to
be over five times greater than the observed amount of visible matter.
The dark matter is traditionally split into hot dark matter and cold
dark matter. Hot dark matter consists of particles that travel with
ultrarelativistic velocities. The best candidate for the identity of hot
dark matter is the neutrino, albeit the observed left-handed neutrinos
with masses of few electron volts cannot constitute the bulk of dark
matter. A right-handed neutrino could be a viable candidate for the
role of dark matter, but such particles have not been detected so far.
Cold dark matter is composed of massive slowly moving and weakly
interacting particles. A number of such particles arise in particle
physics models beyond the standard model \cite{Bertone:2004pz}.
These candidates have been studied extensively with results that
are in agreement with experiments. However, these particles have not
been detected so far. Moreover, the discovery of the accelerated
expansion of the Universe \cite{Riess:1998cb} calls for another new form
of invisible energy, known as dark energy. The dark energy provides
most of the energy density in the Universe and it has to provide
negative pressure. There are many proposals considering the explanation
of dark energy \cite{Copeland:2006wr}, but no compelling candidate.

The alternative possibility for explaining the phenomena that are
attributed to dark energy and dark matter is to revise the
theory of gravity. There exist several kinds of modified or alternative
gravitational theories, e.g., (Brans-Dicke) scalar tensor theories,
Gauss-Bonnet gravities, $f(R)$ gravity, brane world models, conformal
gravity, Poincar\'e gauge theories and many more. In this letter, we
concentrate on TeVeS alone.  TeVeS is a highly interesting theory of
gravity, since it is a relativistic theory, it obeys the Einstein
equivalence principle and produces the MOND phenomenology. On the other
hand, in TeVeS, the gravitational vector and scalar fields are coupled
to the metric of spacetime in a nonminimal way, which means that the
local dynamics of the relativistic theory is involved and rich. The
propagation of perturbations in the linearized theory has been studied
in \cite{Sagi:2010ei}. We study the full nonlinear theory using the
Arnowitt-Deser-Misner (ADM) decomposition of the gravitational field
\cite{Arnowitt:1962hi} and the Hamiltonian formalism. We show that
TeVeS is free from ghosts, which is a necessary condition for the
consistency of the theory.
Ghosts are notorious for causing instability in several theories, for
example, in the renormalizable Weyl-like theories of gravity
\cite{Boulware:1984,Kluson:2013hza}.

\section{Fundamentals of TeVeS}
TeVeS contains extra gravitational degrees of freedom, which are carried
by a vector field $A_\mu$ and a scalar field $\phi$.
We emphasize that TeVeS involves two frames: the Bekenstein frame for
the gravitational fields and a physical frame for the matter fields.
The Bekenstein frame has the metric $\tg_{\mu\nu}$ with the connection
$\tnabla_\mu$. The action for all matter fields is written using
a physical metric $g_{\mu\nu}$ with the connection $\nabla_\mu$, which
is related to the three gravitational fields $\tg_{\mu\nu}$, $A_\mu$ and
$\phi$ as
\begin{equation}\label{g_munu}
g_{\mu\nu}=e^{-2\phi}\tg_{\mu\nu}-2\sinh(2\phi)A_\mu A_\nu .
\end{equation}
The fact that all matter fields couple to the physical metric means that
the Einstein equivalence principle is obeyed.
The vector field is required to be timelike and normalized with respect
to the Bekenstein metric,
\begin{equation}\label{constraA}
A_\mu A^\mu=-1,
\end{equation}
where the covariant index is raised with the Bekenstein metric,
$A^\mu=\tg^{\mu\nu}A_\nu$.
The inverse of the physical metric is obtained as
\begin{equation}\label{g^munu}
g^{\mu\nu}=e^{2\phi}\tg^{\mu\nu}+\frac{2\sinh (2\phi)e^{2\phi}}
{e^{2\phi}-2\sinh (2\phi)(A_\mu A^\mu+1)}A^\mu A^\nu.
\end{equation}

The action of the theory,
\begin{equation}
S=S_{\tg}+S_A+S_\phi+S_m,
\end{equation}
consists of the actions for the metric $\tg_{\mu\nu}$, the vector field
$A_\mu$, the scalar field $\phi$ and matter, respectively.
The action for $\tg_{\mu\nu}$ is defined as the standard
Einstein-Hilbert action,
\begin{equation}\label{Stg}
S_{\tg}=\frac{1}{16\pi G}\int_\mM d^4x\sqrt{-\tg}\tilde{R}
+\frac{1}{8\pi G}\oint_{\partial\mM}d^3x\sqrt{|\gamma|}\tilde{K} ,
\end{equation}
where $G$ is the bare gravitational constant, $\tg=\det \tg_{\mu\nu}$,
and $\tilde{R}$ is the scalar curvature defined by the connection
$\tilde{\nabla}$ which is compatible with the metric $\tg_{\mu\nu}$.
The surface integral over the boundary $\partial\mM$ of the spacetime
$\mM$ is included so that only the variation of the metric
$\delta\tg_{\mu\nu}$ (and not its derivatives) needs to be imposed to
vanish on the boundary, when obtaining the Einstein field equations for
$\tg_{\mu\nu}$. In the surface term, $\gamma$ is the determinant of the
induced metric on $\partial\mM$ and $\tilde{K}$ is the trace of the
extrinsic curvature of $\partial\mM$.

The action for the vector field $A_\mu$ is given by
\begin{equation}\label{SA}
S_A=- \frac{1}{32\pi G}\int_\mM d^4x\sqrt{-\tg}
[\kappa F_{\mu\nu}F^{\mu\nu}-2\lambda (A_\mu A^\mu+1)] ,
\end{equation}
where $F_{\mu\nu}=\tilde\nabla_\mu A_\nu-\tilde\nabla_\nu A_\mu=
\partial_\mu A_\nu-\partial_\nu A_\mu$, $\kappa$ is a dimensionless
constant and $\lambda$ is a Lagrange multiplier ensuring
(\ref{constraA}). The original action (\ref{SA}) has since been extended
with three extra terms which are quadratic in $\tilde{\nabla}_\mu
A_\nu$, see \cite{Skordis:2008}, in order to cure certain dynamical
problems, see e.g. \cite{Sagi:2010ei}. Here we consider the original
TeVeS for simplicity. A detailed Hamiltonian analysis of the
extended TeVeS model will be presented in a future communication.

The action for the scalar field $\phi$ is given by
\begin{equation}\label{Sphi}
S_\phi=-\frac{1}{16\pi G}\int_\mM d^4x \sqrt{-\tg} \left[ \mu
\hg^{\mu\nu}\tnabla_\mu \phi\tnabla_\nu \phi+V(\mu) \right],
\end{equation}
where $\mu$ is a non-dynamical dimensionless scalar field and $\hg$ is
a new metric defined as
\begin{equation}
\hg^{\mu\nu}=\tg^{\mu\nu}-A^\mu A^\nu .
\end{equation}
The potential term $V(\mu)$ is an arbitrary function that typically
depends on a scale. The metric $\hg^{\mu\nu}$ is used in the
scalar field action, rather than $\tg^{\mu\nu}$, in order to avoid
superluminal propagation of perturbations. For the same purpose, we
assume that $\phi>0$ \cite{Bekenstein:2004ne}.

All matter fields, denoted generically by $\chi^A$, are coupled to the
physical metric $g_{\mu\nu}$ so that their action has the form
\begin{equation}
S_m=\int_\mM d^4x\sqrt{-g}\mL[g,\chi^A,\nabla \chi^A].
\end{equation}
For simplicity, we will consider a scalar matter field $\chi$ with
the action
\begin{equation}\label{Schi}
S_m=-\int_\mM d^4x\sqrt{-g}\left[ \frac{1}{2}g^{\mu\nu}
\partial_\mu\chi\partial_\nu\chi+\mV(\chi) \right].
\end{equation}
The determinant of the physical metric $g$ is related to the
determinant of the Bekenstein metric $\tg$ and the fields $\phi$ and
$A_\mu$ as
\begin{equation}\label{g}
g=e^{-4\phi}\left[ 1-(1-e^{-4\phi})(A_\mu A^\mu+1) \right] \tg.
\end{equation}

\section{Hamiltonian structure of TeVeS}
The spacetime is assumed to admit a foliation into a union of
nonintersecting spacelike hypersurfaces
$\Sigma_t$, which are parameterized by the time $t$.
The Bekenstein metric $\tg_{\mu\nu}$ induces a metric $h_{\mu\nu}$ on
$\Sigma_t$, which is defined as
\begin{equation}
h_{\mu\nu}=\tg_{\mu\nu}+n_\mu n_\nu ,
\end{equation}
where $n_\mu$ is the future-directed unit normal to $\Sigma_t$.
The unit normal can be written in terms of the lapse function $N$ and
the shift vector $N^i$ ($i=1,2,3$) as
\begin{equation}
n_\mu=-\nabla_\mu t=(-N,0,0,0),\
n^\mu=\left( \frac{1}{N},-\frac{N^i}{N} \right).
\end{equation}
Consequently, the metric $\tg_{\mu\nu}$ is decomposed in terms of the
ADM variables $h_{ij}$, $N$ and $N^i$ as
\begin{equation}
\tg_{00}=-N^2+N^i h_{ij}N^j , \ \tg_{0i}=h_{ij}N^j ,
\ \tg_{ij}=h_{ij} .
\end{equation}
The determinant of the metric $\tg_{\mu\nu}$ is written as
\begin{equation}\label{sqrt-tildeg}
\sqrt{-\tg}=N\sqrt{h},\  h=\det h_{ij}.
\end{equation}
The vector field $A_\mu$ is decomposed into components tangent and
normal to $\Sigma_t$ as
\begin{equation}
{}_\perp A_\mu=h_\mu^{ \ \nu}A_\nu ,\  A_{\bn}=n^\mu A_\mu ,
\end{equation}
respectively, where $h_\mu^{\phantom\mu\nu}=h_{\mu\rho}\tg^{\rho\nu}
=\delta_\mu^{\phantom\mu\nu}+n_\mu n^\nu$ is the projection operator
onto $\Sigma_t$. That is the components of the vector field are
expressed as $A_0=NA_\bn+N^iA_i$ and $A_i={}_\perp A_i$.

In the Hamiltonian formulation of TeVeS, the canonical momenta conjugate
to $h_{ij}$, $N$, $N^i$, $A_\bn$, $A_i$, $\lambda$, $\phi$, $\mu$ and
$\chi$ are denoted by $\pi^{ij}$, $\pi_N$, $\pi_i$, $p_{\bn}$, $p^i$,
$p_\lambda$, $p_\phi$, $p_\mu$ and $p_\chi$, respectively.
Since the action is independent of the time derivatives of $N$, $N^i$,
$\lambda$, $A_\bn$ and $\mu$, their canonically conjugated
momenta are the primary constraints:
\begin{equation}\label{pc}
\pi_N\approx 0 ,\  \pi_i\approx 0 ,\  p_\lambda\approx 0
,\  p_{\bn}\approx 0 ,\  p_\mu\approx 0.
\end{equation}
We obtain the total Hamiltonian in the form (with all
primary constraints included through Lagrange multipliers)
\begin{eqnarray}\label{H}
H=\int_{\Sigma_t}d^3x (N\mH_T+N^i\mH_i+v_N\pi_N+v^i\pi_i
\nonumber\\
+v_\lambda p_\lambda +v_{\bn}p_{\bn} +v_\mu p_\mu)
+H_\mathrm{surf},
\end{eqnarray}
where $\mH_T$ is the Hamiltonian constraint, $\mH_i$ is the momentum
constraint, and $H_\mathrm{surf}$ is the surface contribution.
The momentum constraint has the form
\begin{eqnarray}
\mH_i= -2h_{ij}D_k\pi^{jk}
-A_i\partial_jp^j+(\partial_iA_j-\partial_jA_i)p^j \nonumber \\
+\ \partial_i\phi p_\phi+\partial_i\chi p_\chi \approx 0,
\end{eqnarray}
where $D_k$ is the covariant derivative compatible with the metric
$h_{ij}$.
The momentum constraint defines the generator of the time-dependent
spatial diffeomorphisms for the dynamical variables on $\Sigma_t$.
The Hamiltonian constraint is responsible for the time evolution of the
canonical variables. It consists of the contributions of the tensor,
vector, scalar and matter fields,
\begin{equation}\label{H_T}
\mH_T=\mH_T^\mathrm{GR}+\mH_T^A+\mH_T^\phi+\mH_T^\chi
\approx 0,
\end{equation}
respectively.
The tensor contribution is similar to GR,
\begin{equation}
\mH_T^\mathrm{GR}=\frac{16\pi G}{\sqrt{h}}\pi^{ij}\mG_{ijkl}\pi^{kl}
-\frac{\sqrt{h}}{16\pi G}\sR,
\end{equation}
where
\begin{equation}
\mG_{ijkl}=\frac{1}{2}(h_{ik}h_{jl}+h_{il}h_{jk}) -\frac{1}{2}
h_{ij}h_{kl}
\end{equation}
and $\sR$ is the scalar curvature defined by the covariant derivative
$D_i$. The contributions of the vector and scalar fields are
\begin{eqnarray}
\mH_T^A&=&\frac{4\pi G}{\kappa\sqrt{h}} p^ih_{ij}p^j
-A_{\bn}D_ip^i \nonumber \\
&+&\frac{\kappa}{32\pi G}\sqrt{h}h^{ik}h^{jl}
(D_iA_j-D_jA_i)(D_kA_l-D_lA_k) \nonumber\\
&+&\frac{\lambda}{16\pi G}\sqrt{h}( A_iA^i-A_\bn^2+1)
\end{eqnarray}
and
\begin{eqnarray}
\mH_T^\phi&=& \frac{4\pi G}{\sqrt{h}\mu (1+A_\bn^2)}p_\phi^2
+\frac{A_{\bn}}{(1+A_\bn^2)} p_\phi A^i\partial_i\phi \nonumber \\
&-& \frac{\mu\sqrt{h}}{16\pi G(1+A_\bn^2)}(A^i\partial_i\phi)^2
\nonumber \\
&+&\frac{1}{16\pi G}\mu\sqrt{h}h^{ij}\partial_i\phi\partial_j\phi
+V(\mu).
\end{eqnarray}
The contribution of the matter field is the most interesting one,
since it contains the contribution of the nonminimal coupling between
the gravitational tensor, vector and scalar fields due to
(\ref{g_munu}).
The matter part of the Hamiltonian constraint is given as
\begin{eqnarray}
\mH_T^{\chi}&=&\frac{\sqrt{1-(1-e^{-4\phi})\mG_\lambda}}
{2\sqrt{h} (e^{-4\phi}-(1-e^{-4\phi})A_iA^i)}p_\chi^2 \nonumber\\
&-&\frac{(1-e^{-4\phi})A_{\bn}}{e^{-4\phi}-(1-e^{-4\phi})A_iA^i}
A^i\partial_i\chi p_\chi \nonumber\\
&+&\sqrt{h\left(1-(1-e^{-4\phi})\mG_\lambda\right)}\times \nonumber\\
&\times& \left[ \frac{1-e^{-4\phi}}{2(e^{-4\phi}-(1-e^{-4\phi})A_iA^i)}
(A^i\partial_i\chi)^2 \right.\nonumber\\
&+&\left. \frac{1}{2}h^{ij}\partial_i\chi\partial_j\chi
+e^{-2\phi}\mV(\chi) \right]. \label{Hchi_T}
\end{eqnarray}
Three more constraints are required in order to ensure that the
primary constraints (\ref{pc}) are preserved in time,
\begin{eqnarray}
\mG_\lambda &=& A_iA^i-A_{\bn}^2+1 \approx 0,\label{G_lambda}\\
\mG_{\bn} &=& D_ip^i +\frac{\lambda\sqrt{h}}{8\pi G}A_{\bn}+\dots
\approx 0, \label{G_n}\\
%+\frac{8\pi GA_\bn}{\sqrt{h}\mu(1+A_\bn^2)^2}p_\phi^2
%-\frac{1}{1+A_\bn^2}p_\phi A^i\partial_i\phi \nonumber \\
%&+& 2\frac{A_\bn^2}{(1+A_\bn^2)^2}p_\phi A^i\partial_i\phi
%-\frac{\sqrt{h}\mu A_\bn}{8\pi G(1+A_\bn^2)^2} (A^i\partial_i\phi)^2
%\nonumber \\
%&-& \frac{\sinh(2\phi)A_{\bn}}{\sqrt{\mh}(e^{2\phi}
%-2\sinh(2\phi)A_{\bn}^2)^2}p_\chi^2
%+\frac{2\sinh(2\phi)(e^{2\phi}+2\sinh(2\phi)A_\bn^2)}
%{(e^{2\phi}-2\sinh(2\phi)A_{\bn}^2)^2}A^i \partial_i\chi p_\chi
%\nonumber \\
%&-& \frac{6\sqrt{\mh}e^{2\phi}\sinh^2(2\phi)A_{\bn}}
%{(e^{2\phi}-2\sinh(2\phi)A_{\bn}^2)^2}(A^i\partial_i\chi)^2
%-\frac{\sqrt{\mh}e^{2\phi}\sinh(2\phi)A_{\bn}}{e^{2\phi}-2\sinh(2\phi)
%A_{\bn}^2}h^{ij}\partial_i\chi\partial_j\chi
%\nonumber \\
%&-& \frac{2\sqrt{\mh}\sinh(2\phi)A_{\bn}}
%{e^{2\phi}-2\sinh(2\phi)A_{\bn}^2}\mV(\chi) \approx 0 ,\label{G_n}\\
\mG_\mu &=& \frac{4\pi G}{\sqrt{h}\mu^2(1+A_\bn^2)} p_\phi^2
+\frac{\sqrt{h}}{16\pi G(1+A_\bn^2)}(A^i\partial_ip_\phi)^2 \nonumber \\
&-& \sqrt{h}h^{ij}\partial_i\phi\partial_j\phi-\sqrt{h}\frac{\delta
V(\mu)} {\delta\mu} \approx 0 ,
\end{eqnarray}
where the constraint (\ref{G_n}) has a complicated form, involving
all the dynamical variables, and it has been omitted.
The constraint (\ref{G_lambda}) was introduced already in
(\ref{Hchi_T}).

Note that the spatial part of the physical metric (\ref{g_munu}), namely
$g_{ij}=e^{-2\phi}h_{ij}-2\sinh(2\phi)A_iA_j$,
changes its signature when the scalar field $\phi$ becomes larger than
$\frac{1}{4}\ln\left( 1+(A_iA^i)^{-1} \right)$.
This can be seen in the determinant
\begin{equation}\label{detg_ij}
\det(g_{ij})=e^{-2\phi}\left( e^{-4\phi}-(1-e^{-4\phi})A_iA^i \right) h.
\end{equation}
The change of signature is reflected in the matter part of the
Hamiltonian constraint (\ref{Hchi_T}), where the denominator of the
first three terms contains the same expression as (\ref{detg_ij}),
$e^{-4\phi}-(1-e^{-4\phi})A_iA^i$.
These terms diverge at $\phi=\frac{1}{4}\ln\left( 1+(A_iA^i)^{-1}
\right)$ and change their signs thereafter. In particular, the kinetic
term $p_\chi^2$ becomes negative if $\phi$ is allowed to pass this
point. In order to obtain a well-defined Hamiltonian formulation of
matter in TeVeS, we require that the hypersurfaces $\Sigma_t$ are
spacelike in the physical frame. Combined with the requirement of no
superluminal propagation of perturbations, $\phi>0$, we obtain the
restriction
\begin{equation}\label{phiregion}
0<\phi<\frac{1}{4}\ln\left( 1+\frac{1}{A_iA^i} \right).
\end{equation}
When the unit timelike vector field $A_\mu$ is dominated by the
component $A_\bn$, we have a weak spatial vector $A_i$,
$0\le A_iA^i\ll1$, and hence the permitted region (\ref{phiregion}) for
$\phi$ is large. Conversely, if $A_iA^i\gg1$, the permissible region for
$\phi$ is narrow, with an upper limit of order $1/(4A_iA^i)$.

The first class constraints $\pi_N,\pi_i,\mH_T,\mH_i$ are associated
with the invariance of the original theory under four-dimensional
diffeomorphisms.
The second class constraints $p_\lambda,p_{\bn},p_\mu,\mG_\lambda,
\mG_{\bn},\mG_\mu$ can be used to express the variables
$\lambda,A_\bn,\mu$ in terms of the gravitational variables $h_{ij}$,
$A_i$, $\phi$ and the matter fields.

The surface term $H_\mathrm{surf}$ in the Hamiltonian (\ref{H}) defines
the total gravitational energy in space. The physical Hamiltonian is
given by $H_\mathrm{phys}=H-H_\mathrm{b}$, where $H_\mathrm{b}$ is the
Hamiltonian for a given reference background. We define the total energy
associated with the time translation along $t^\mu=Nn^\mu+N^\mu$ for any
given solution of the equations of motion as the value of the physical
Hamiltonian when all the constraints are satisfied. For a stationary
background, we obtain the total gravitational energy as
\begin{eqnarray}\label{E}
E=-\oint_{\partial\Sigma_t}d^2x\left( \frac{1}{8\pi G}N\sqrt{\sigma}
\left( \Kt-\Kt_\mathrm{b} \right)\right.\nonumber \\
\left.  -2N^i h_{ij}r_k\pi^{jk} -r_ip^i(NA_\bn+N^jA_j)  \right),
\end{eqnarray}
where $\sigma$, $\Kt$ and $r_i$ are the determinant of the induced
metric, the trace of the extrinsic curvature and the unit normal for the
boundary of $\Sigma_t$, respectively, $\Kt_\mathrm{b}$ is the trace of
the extrinsic curvature of the boundary on the reference background, and
$A_\bn$ is given by the constraint $\mG_\lambda=0$ as
$A_\bn=\pm\sqrt{A_iA^i+1}$. The expression for the total energy
(\ref{E}) of TeVeS differs from that of GR in two ways: the metric on
$\Sigma_t$ is induced by $\tg_{\mu\nu}$ (not by $g_{\mu\nu}$) and the
contribution of the vector field is included, namely the last term
$\oint_{\partial\Sigma_t}d^2x r_ip^iA_0$. This generic expression for
the total energy can be used to obtain the total energy with respect to
different kinds of backgrounds, as in GR \cite{Hawking:1995fd}.

For an asymptotically flat spacetime, the expression (\ref{E}) becomes
the ADM energy of TeVeS. Recall that in GR the ADM energy satisfies the
positive energy theorem \cite{Schoen:1979zz,Witten:1981cmp}. We have not
proven the positivity of the total energy (\ref{E}) for an arbitrary
isolated system, albeit we do expect that the positive energy theorem
will hold in TeVeS. The ADM energy of the flat Minkowski spacetime is
zero by definition. As a nontrivial example, we consider the spherically
symmetric solution of the field equations of TeVeS with a vanishing
radial vector component ($A^r=0$) which was obtained in
\cite{Giannios:2005es} using isotropic spherical coordinates as
\begin{equation}\label{SpherSol}
\tg_{tt}=-\left( \frac{r-r_c}{r+r_c} \right)^{r_g/2r_c},\quad
\tg_{rr}=\frac{(r^2-r_c^2)^2}{r^4}\left( \frac{r-r_c}{r+r_c}
\right)^{-r_g/2r_c},
\end{equation}
where the characteristic radius is defined as
\begin{equation}\label{r_c}
r_c=\frac{r_g}{4}\sqrt{1+\frac{k}{\pi}\left( \frac{Gm_s}{r_g} \right)^2
-\frac{\kappa}{2}},
\end{equation}
and where  the ``scalar mass'' $m_s$ and the gravitational radius $r_g$
are related to the total gravitational mass \cite{Bekenstein:2004ne},
and $k$ is a dimensionless constant involved in the definition of the
potential in the action (\ref{Sphi}). We obtain the ADM energy of the
solution (\ref{SpherSol}) as
\begin{equation}
E_{\mathrm{ADM}}
% =\frac{1}{16\pi G}\oint_{S}d^2x\sqrt{\sigma}
% \left( D^ih_{ij}-D_jh \right)r^j
=-\frac{1}{2G}\lim_{r\rightarrow\infty}r^2
\frac{\partial h_{rr}}{\partial r}
% =\frac{r_g}{2G}\lim_{r\rightarrow\infty}\left( 1-\frac{4r_c^2}{r_gr}
% \right)\left( 1-\frac{r_c^2}{r^2} \right)\left( \frac{r-r_c}{r+r_c}
% \right)^{-r_g/2r_c}
=\frac{r_g}{2G}.
\end{equation}
The ADM energy depends on $r_g$ rather than on the characteristic radius
(\ref{r_c}) of the solution. Identifying the ADM energy as the
gravitational mass $m$ of an isolated spherical matter distribution,
one obtains $r_g=2Gm$.

\section{Discussion}
We have uncovered the Hamiltonian structure of the original version of
the TeVeS theory of gravity \cite{Bekenstein:2004ne}. TeVeS is shown to
contain six local gravitational degrees of freedom: two in the usual
spin-2 graviton, three in the unit timelike vector field, and one in the
scalar field. This is consistent with the previous knowledge on the
theory. However, there is an important detail regarding the linearized
theory. When we consider the lowest order perturbations in the absence
of matter, with the background given as $\tg_{\mu\nu}=\eta_{\mu\nu}
=\mathrm{diag}(-1,1,1,1)$, $A^\mu=(1,0,0,0)$ and
$\phi=\phi_c=\mathrm{constant}$, the tensor perturbation has two
traceless-transverse modes and the vector perturbation has two
transverse modes, which all propagate at the uniform speed
$e^{-2\phi_c}$ \cite{Sagi:2010ei}. The scalar perturbation is a trace
mode. The third degree of freedom which is associated with the  vector
field in the full nonlinear theory does not appear in the linearized
theory. It appears, however, as an extra trace mode when the action of
the vector field (\ref{SA}) is generalized (see \cite{Sagi:2010ei}).
Hence, in the case of the original version of TeVeS, the lowest order
linearized theory lacks one degree of freedom.

The nonminimal coupling of the vector and scalar fields to the
Bekenstein metric was found to be intricate, yet well defined. In the
present Hamiltonian formulation, the nonminimal coupling is contained in
the matter part (\ref{Hchi_T}) of the Hamiltonian constraint
(\ref{H_T}). The kinetic terms in the Hamiltonian constraint are
positive definite, assuming (\ref{phiregion}) is satisfied, and hence
there is no sign of ghost instability in TeVeS.
This offers further support for the theoretical soundness of TeVeS.
Complemented by the remarkable success of TeVeS in explaining the
observed discrepancy between the dynamics and the distribution of the
visible matter in galaxies, we can conclude that TeVeS is a highly
interesting proposal for the extension of GR.
There are further challenges and prospects.

It appears that some dark matter is still required in TeVeS, since
otherwise TeVeS is unable to explain certain observations on galaxy
clusters, gravitational lensing and the cosmic microwave background
radiation \cite{Feix:2010pn,Clifton:2011jh}.
It has been hypothesized that the required nonluminous matter could be
composed of massive (sterile) neutrinos.

It is known that the Einstein-Hilbert action of GR (including the
cosmological constant) is generated at one-loop order in any quantum
field theory, when the geometry of the background spacetime is not
fixed in the beginning
\cite{Sakharov:1968,Adler:1982,Zee:1983,Smilga:1984}.
This includes renormalizable higher-order derivative theories of
gravity, such as Weyl gravity, where GR is induced at long distances.
Obtaining TeVeS via such an induced mechanism is difficult, since matter
couples to the physical metric. Hence the induced GR term is the
physical curvature $R$, not the curvature $\tilde R$ defined by the
Bekenstein metric. We speculate that the gravitational vector and
scalar fields need to be present from the beginning and with
nonminimal coupling to the background. Conceivably, in such a setting,
the one-loop quantum corrections could generate the curvature part
(\ref{Stg}) of the TeVeS action. How else could the nonminimal coupling
emerge?

These considerations do not address the quantum aspects of gravity
itself. The quantization of TeVeS is indeed expected to be just as
challenging as the quantization of GR.

\subsection*{Acknowledgements}
We are indebted to J. Bekenstein and M. Milgrom for useful discussions
and suggestions.
The support of the Academy of Finland under Project Nos. 136539
and 272919, as well as of the Magnus Ehrnrooth Foundation, is gratefully
acknowledged.
The work of J.K. was supported by the Grant Agency of the Czech Republic
under Grant P201/12/G028.
The work of M.O. was also supported by the Jenny and Antti Wihuri
Foundation.

\end{document}